\documentstyle[prd,aps,twocolumn,amstex,floats]{revtex}
\input epsf
\newcommand{\ifm}[1]{\relax\ifmmode #1\else $#1$\fi}
\newcommand{\etal}{{\it et al.}}
\newcommand{\Dzero}{D\O}
\newcommand{\cdf}{CDF Collaboration}
\newcommand{\dzero}{\Dzero\ Collaboration}
\newcommand{\PRL}{Phys. Rev. Lett.}
\newcommand{\PL}{Phys. Lett.}
\newcommand{\PR}{Phys. Rev.}
\newcommand{\NP}{Nucl. Phys.}

\begin{document}
\pagestyle{myheadings}
\onecolumn

\title{Combination of CDF and D\O\ Results on $W$ Boson Mass and Width}

%
\def\r#1{\ignorespaces $^{#1}$}
\author{                                                                      
V.M.~Abazov,$^{36,\S}$                                                           
B.~Abbott,$^{94,\S}$                                                             
A.~Abdesselam,$^{13,\S}$                                                         
M.~Abolins,$^{82,\S}$                                                            
V.~Abramov,$^{39,\S}$                                                            
B.S.~Acharya,$^{20,\S}$                                                          
D.~Acosta,\r {57,\dag}
D.L.~Adams,$^{90,\S}$                                                            
M.~Adams,$^{62,\S}$                                                              
 T.~Affolder,\r {55,\dag}
% ...
\vskip 0.30cm                                                                 
\centerline{($^{\dag}$CDF Collaboration)}      
\centerline{($^{\S}$D\O\ Collaboration)}     
\centerline{($^{\ddag}$Tevatron Electroweak Working Group)}      
\vskip 0.30cm
}
\address{                                                                     
\centerline{$^{1}$Universidad de Buenos Aires, Buenos Aires, Argentina}       
\centerline{$^{2}$LAFEX, Centro Brasileiro de Pesquisas F{\'\i}sicas,    Rio de Janeiro, Brazil}                                     
\centerline{$^{3}$Universidade do Estado do Rio de Janeiro, Rio de Janeiro, Brazil} 
\centerline{$^{4}${Institute of Particle Physics, McGill University, Montreal, Canada H3A 2T8, and University of Toronto, Toronto M5S 1A7, Canada}} 
\centerline{$^{5}$Institute of High Energy Physics, Beijing,  People's Republic of China}                                 
\centerline{$^{6}${Institute of Physics, Academia Sinica, Taipei, Taiwan 11529, Republic of China}} 
\centerline{$^{7}$Universidad de los Andes, Bogot\'{a}, Colombia}             
\centerline{$^{8}$Institute of Physics, Academy of Sciences, Center  for Particle Physics, Prague, Czech Republic}               
\centerline{$^{9}$Charles University, Center for Particle Physics,   Prague, Czech Republic}                                     
\centerline{$^{10}$Universidad San Francisco de Quito, Quito, Ecuador}         
\centerline{$^{11}$Laboratoire de Physique Subatomique et de Cosmologie,  IN2P3-CNRS, Universite de Grenoble 1, Grenoble, France}     
\centerline{$^{12}$CPPM, IN2P3-CNRS, Universit\'e de la M\'editerran\'ee,  Marseille, France}                                          
\centerline{$^{13}$Laboratoire de l'Acc\'el\'erateur Lin\'eaire, IN2P3-CNRS, Orsay, France}                                  
\centerline{$^{14}$LPNHE, Universit\'es Paris VI and VII, IN2P3-CNRS,  Paris, France}                                              
\centerline{$^{15}$DAPNIA/Service de Physique des Particules, CEA, Saclay,  France}   
\centerline{$^{16}$Universit{\"a}t Freiburg, Physikalisches Institut,  Freiburg, Germany}                                          
\centerline{$^{17}${Institut f\"{u}r Experimentelle Kernphysik, Universit\"{a}t Karlsruhe, 76128 Karlsruhe, Germany}} 
\centerline{$^{18}$Panjab University, Chandigarh, India}     
\centerline{$^{19}$Delhi University, Delhi, India}                            
\centerline{$^{20}$Tata Institute of Fundamental Research, Mumbai, India}     
\centerline{$^{21}${University College Dublin, Belfield, Dublin 4, Ireland}} 
\centerline{$^{22}${Istituto Nazionale di Fisica Nucleare, University of Bologna,I-40127 Bologna, Italy}} 
\centerline{$^{23}${Laboratori Nazionali di Frascati, Istituto Nazionale di Fisica Nucleare, I-00044 Frascati, Italy}} 
\centerline{$^{24}${Universita di Padova, Istituto Nazionale di Fisica Nucleare, Sezione di Padova, I-35131 Padova, Italy}} 
\centerline{$^{25}${Istituto Nazionale di Fisica Nucleare, University and Scuola   Normale Superiore of Pisa, I-56100 Pisa, Italy}} 
\centerline{$^{26}${Instituto Nazionale de Fisica Nucleare, Sezione di Roma, University di Roma I, ``La Sapienza," I-00185 Roma, Italy}}
\centerline{$^{27}${Istituto Nazionale di Fisica Nucleare, University of Trieste/Udine, Italy}} 
\centerline{$^{28}${Hiroshima University, Higashi-Hiroshima 724, Japan}} 
\centerline{$^{29}${University of Tsukuba, Tsukuba, Ibaraki 305, Japan}}
\centerline{$^{30}${Osaka City University, Osaka 588, Japan}} 
\centerline{$^{31}${Waseda University, Tokyo 169, Japan}} 
\centerline{$^{32}${Center for High Energy Physics: Kyungpook National University, Taegu 702-701; Seoul National University, Seoul 151-742;}} 
\centerline{ {and SungKyunKwan University, Suwon 440-746; Korea}} 
\centerline{$^{33}$CINVESTAV, Mexico City, Mexico}                            
\centerline{$^{34}$FOM-Institute NIKHEF and University of  Amsterdam/NIKHEF, Amsterdam, The Netherlands}               
\centerline{$^{35}$University of Nijmegen/NIKHEF, Nijmegen, The  Netherlands}                                                
\centerline{$^{36}$Joint Institute for Nuclear Research, Dubna, Russia}       
\centerline{$^{37}${Institution for Theoretical and Experimental Physics, ITEP, Moscow 117259, Russia}} 
\centerline{$^{38}$Moscow State University, Moscow, Russia}                   
\centerline{$^{39}$Institute for High Energy Physics, Protvino, Russia}       
\centerline{$^{40}${Instituto de Fisica de Cantabria, CSIC-University of Cantabria, 39005 Santander, Spain}} 
\centerline{$^{41}${University of Geneva, CH-1211 Geneva 4, Switzerland}} 
\centerline{$^{42}${Glasgow University, Glasgow G12 8QQ, United Kingdom}}
\centerline{$^{43}$Lancaster University, Lancaster, United Kingdom}           
\centerline{$^{44}$Imperial College, London, United Kingdom}                  
\centerline{$^{45}${University College London, London WC1E 6BT, United Kingdom}} 
\centerline{$^{46}${University of Oxford, Oxford OX1 3RH, United Kingdom}} 
\centerline{$^{47}$University of Arizona, Tucson, Arizona 85721}  
\centerline{$^{48}$Lawrence Berkeley National Laboratory and University of  California, Berkeley, California 94720}              
\centerline{$^{49}${Ernest Orlando Lawrence Berkeley National Laboratory, Berkeley, California 94720}} 
\centerline{$^{50}${University of California at Davis, Davis, California  95616}} 
\centerline{$^{51}$California State University, Fresno, California 93740}     
\centerline{$^{52}$University of California, Irvine, California 92697}        
\centerline{$^{53}${University of California at Los Angeles, Los Angeles, California  90024}} 
\centerline{$^{54}$University of California, Riverside, California 92521}     
\centerline{$^{55}${University of California at Santa Barbara, Santa Barbara, California 93106}}  
\centerline{$^{56}${Yale University, New Haven, Connecticut 06520}} 
\centerline{$^{57}${University of Florida, Gainesville, Florida  32611}} 
\centerline{$^{58}$Florida State University, Tallahassee, Florida 32306}      
\centerline{$^{59}${Argonne National Laboratory, Argonne, Illinois 60439}} 
\centerline{$^{60}$Fermi National Accelerator Laboratory, Batavia,  Illinois 60510}                  
\centerline{$^{61}${Enrico Fermi Institute, University of Chicago, Chicago, Illinois 60637}} 
\centerline{$^{62}$University of Illinois at Chicago, Chicago,   Illinois 60607}                   
\centerline{$^{63}$Northern Illinois University, DeKalb, Illinois 60115}      
\centerline{$^{64}$Northwestern University, Evanston, Illinois 60208}         
\centerline{$^{65}${University of Illinois, Urbana, Illinois 61801}} 
\centerline{$^{66}$Indiana University, Bloomington, Indiana 47405}            
\centerline{$^{67}$University of Notre Dame, Notre Dame, Indiana 46556}       
\centerline{$^{68}${Purdue University, West Lafayette, Indiana 47907} }
\centerline{$^{69}$Iowa State University, Ames, Iowa 50011}                   
\centerline{$^{70}$University of Kansas, Lawrence, Kansas 66045}              
\centerline{$^{71}$Kansas State University, Manhattan, Kansas 66506}          
\centerline{$^{72}$Louisiana Tech University, Ruston, Louisiana 71272}        
\centerline{$^{73}${The Johns Hopkins University, Baltimore, Maryland 21218}} 
\centerline{$^{74}$University of Maryland, College Park, Maryland 20742}      
\centerline{$^{75}$Boston University, Boston, Massachusetts 02215}            
\centerline{$^{76}$Northeastern University, Boston, Massachusetts 02115}      
\centerline{$^{77}${Harvard University, Cambridge, Massachusetts 02138}} 
\centerline{$^{78}${Massachusetts Institute of Technology, Cambridge, Massachusetts  02139}}    
\centerline{$^{79}${Tufts University, Medford, Massachusetts 02155}} 
\centerline{$^{80}${Brandeis University, Waltham, Massachusetts 02254}} 
\centerline{$^{81}${University of Michigan, Ann Arbor, Michigan 48109}} 
\centerline{$^{82}${Michigan State University, East Lansing, Michigan  48824}} 
\centerline{$^{83}$University of Nebraska, Lincoln, Nebraska 68588}           
\centerline{$^{84}${Rutgers University, Piscataway, New Jersey 08855}} 
\centerline{$^{85}${University of New Mexico, Albuquerque, New Mexico 87131}} 
\centerline{$^{86}$Columbia University, New York, New York 10027}             
\centerline{$^{87}${Rockefeller University, New York, New York 10021} }
\centerline{$^{88}$University of Rochester, Rochester, New York 14627}        
\centerline{$^{89}$State University of New York, Stony Brook,   New York 11794}                   
\centerline{$^{90}$Brookhaven National Laboratory, Upton, New York 11973}     
\centerline{$^{91}${Duke University, Durham, North Carolina  27708}} 
\centerline{$^{92}${The Ohio State University, Columbus, Ohio  43210}} 
\centerline{$^{93}$Langston University, Langston, Oklahoma 73050}             
\centerline{$^{94}$University of Oklahoma, Norman, Oklahoma 73019}            
\centerline{$^{95}${University of Pennsylvania, Philadelphia,  Pennsylvania 19104}}    
\centerline{$^{96}${Carnegie Mellon University, Pittsburgh, Pennsylvania  15213}} 
\centerline{$^{97}${University of Pittsburgh, Pittsburgh, Pennsylvania 15260}} 
\centerline{$^{98}$Brown University, Providence, Rhode Island 02912}          
\centerline{$^{99}$University of Texas, Arlington, Texas 76019}               
\centerline{$^{100}${Texas A\&M University, College Station, Texas 77843}} 
\centerline{$^{101}$Rice University, Houston, Texas 77005}                     
\centerline{$^{102}${Texas Tech University, Lubbock, Texas 79409}} 
\centerline{$^{103}$University of Virginia, Charlottesville, Virginia 22901}   
\centerline{$^{104}$University of Washington, Seattle, Washington 98195} 
\centerline{$^{105}${University of Wisconsin, Madison, Wisconsin 53706}} 
\centerline{$^{106}${ High
 Energy Accelerator Research Organization (KEK), Tsukuba, Ibaraki 305, Japan}} 
}                                                                             
\maketitle
\vskip 0.30cm     
\centerline{\today } 
\begin{abstract}
 The results based on 1992-95 data (Run 1)
 from the CDF and D\O\ experiments on the measurements of the
 $W$ boson mass and width are presented, along with the combined results. 
 We report
 a Tevatron collider average $M_W = 80.456 \pm 0.059 $ GeV. We also report the
 Tevatron collider average of the directly measured $W$ boson width 
 $\Gamma_W = 2.115 \pm 0.105 $ GeV. We describe a new joint analysis of
 the direct $W$ mass and width measurements. 
 Assuming the validity of the standard model, we combine the directly measured
 $W$ boson width with the width extracted from the ratio of $W$ and $Z$ boson
 leptonic partial cross sections. This combined result for the Tevatron is 
$\Gamma_W = 2.135 \pm 0.050 $ GeV.  Finally, we use the measurements of the
 direct total 
 $W$ width and the leptonic branching ratio to extract the leptonic
 partial width $\Gamma (W \rightarrow e \nu) = 224 \pm 13$ MeV. 
\end{abstract}

\pacs{PACS numbers: 14.70.Fm, 12.15.Ji, 13.38.Be, 13.85.Qk }


\section{Introduction}
We present new combined
results on the $W$ boson mass and width from the CDF and D\O\ experiments
at the Fermilab Tevatron.
 We document the combination methodologies and summarize the
 results and the various sources of uncertainty, identifying those sources
 that produce correlated uncertainty between the two experiments' 
 results. We also
 present the combination with the UA2 and LEP results. These measurements
 represent some of the main goals
  of the electroweak physics program at the Tevatron
 collider. 

 The $W$ boson mass and width are  important parameters in the electroweak
 gauge sector of the standard model(SM)~\cite{SM}. The $W$ boson, along with 
 the $Z$ boson and the photon, provides a unified description of the
 electroweak interaction as a gauge interaction with the symmetry group
 $SU(2)_L \times U(1)$. If this were an unbroken symmetry, the $W$ and
 $Z$ bosons would be massless. The mass of the $W$ boson and its couplings,
 which determine its width, are therefore of
substantial relevance to tests of the
 structure of the theory and the nature of electroweak symmetry breaking.

 In the SM the $W$ boson mass is related to other parameters, 
 and in the ``on shell'' scheme~\cite{onshell} it can be written as
\begin{equation}
M_W = \left({\pi\alpha\over \sqrt{2} G_F}\right)
^{1\over2}{M_Z\over\sqrt{M_Z^2-M_W^2}\sqrt{1-\Delta r}} \quad ,
\label{eq:mw1}
\end{equation}
where $\alpha$ is the electromagnetic coupling constant and $G_F$ is
 Fermi coupling constant measured in muon decay.
 The electroweak radiative correction
 $\Delta r$ receives calculable
 contributions from loops containing the $t-b$ quarks,
 the Higgs boson (which is the hypothetical agent of electroweak symmetry
 breaking), and any other hypothetical particles such as supersymmetric
 particles coupling to the $W$ boson. Since the top quark mass has been
 measured~\cite{cdfmtop,d0mtop}, the $t-b$ loop correction can be calculated. 
 A precise measurement of the $W$ boson mass therefore constrains the mass
 of the Higgs boson, which has not yet been experimentally observed.  
 Should the Higgs boson be discovered in the future,
 the comparison between its directly measured mass and the indirect 
 constraint will be a very interesting test of the SM. 
In the minimal supersymmetric
extension of the standard model (MSSM), for example,
loop corrections due to supersymmetric particles can contribute 
up to 250~MeV~\cite{susy} to the predicted
$W$ boson mass.

In the SM $W$ bosons decay leptonically: $W \rightarrow
 l \nu$ where $l \in \{e, \mu, \tau \} $, or hadronically: $W \rightarrow q^\prime 
 \bar{q}$, where $q, q^\prime \in \{u, d, c, s, b \} $. The leptonic partial
 width can be calculated~\cite{rosner} :
\begin{equation}
\Gamma ( W \rightarrow e \nu) = \frac{G_F M_W^3}{6\sqrt{2}\pi}\left( 1 + 
 \delta \right) \quad ,
\label{eq:gw1}
\end{equation}
where the SM radiative correction $\delta$ is calculated
 to be less than 0.5\%. Including the QCD radiative 
 corrections for the quark decay channels, the SM prediction for the
 leptonic branching ratio~\cite{PDG2002} is:
\begin{equation}
B ( W \rightarrow e \nu) = \left( \; 3 + 6 \; [ \; 1 + \alpha_s(M_W) /\pi 
 + {\cal O} (\alpha_s^2) \; ]  \; \right) ^{-1} \quad .
\label{eq:bw1}
\end{equation}
Given the precision of these SM calculations, their comparison
 with the measured $W$ boson width provides an important test of the SM.

The precision of the $W$ boson mass and width measurements from the LEP
 experiments (ALEPH, DELPHI, L3 and OPAL) and the Tevatron collider
 experiments (CDF and D\O) is similar, implying that our best knowledge comes
 from the combined results of all these experiments. The measurements
 are quite intricate with many inputs and incorporate constraints from
 data and physics models. In this situation a simple average of
 all measurements, with the  assumption
 that they are completely independent, may be biased in the value or 
 the uncertainty. 

 In this paper we present systematic analyses of the $W$ mass and width
 measurements published by the CDF and D\O\ experiments at Fermilab. Following
 a brief description of the observables in Section~\ref{observables},
 we discuss our methodology and
 calculations. In Section~\ref{wmass}, we consider the $W$ boson mass as
 the parameter of interest and consider all other parameters needed for its
 measurement as external inputs, including the $W$ width.
  We review the uncertainties on these external
 parameters as described in the respective CDF and D\O\ publications, 
 identifying the relevant correlations. This information is used to construct
 the covariance matrix for combining the $W$ mass measurements.

 In Section~\ref{directwwidth}, we perform the same analysis for the
 direct measurement of the $W$ boson width. This is again
 a one-parameter analysis, where in particular the $W$ mass is treated as an
 external input. 

 In Section~\ref{joint}, we present a new methodology for treating the
 $W$ boson mass and width simultaneously in a two-parameter analysis. This
 method departs from the previously published results in that no external
 information is used for either of these parameters. This has two significant
 advantages over the one-parameter analyses that are usually performed:
 (i) theoretical model dependence is reduced, making the results more 
 meaningful and easier to interpret, and (ii) correlation with other
 methods of measuring the $W$ mass and width is reduced, making subsequent
 comparisons and combinations more powerful. We also 
 demonstrate that the joint two-parameter 
 analysis results in no loss of precision, compared to the separate 
 one-parameter analyses of the $W$ mass and width. 

 In Section~\ref{indirectwwidth}, we review the analyses of the ratio ($R$)
 of $W$ and $Z$ boson cross sections in the leptonic channels, as published
 by the CDF and D\O\ collaborations. With some SM assumptions 
 and measured inputs, this ratio can be converted into a measurement of
 the leptonic branching ratio of the $W$ boson, and further into a measurement
 of the $W$ boson width. We present an analysis of the correlated uncertainties
 and the external inputs used to extract the $W$ width. Assumptions made
 in the extraction of the $W$ width from $R$ are compared and contrasted with
 the direct lineshape measurement in Section~\ref{observables}.  

 We conclude the paper with Section~\ref{suggestions},
 discussing future implementations
 of the methodologies presented here. We suggest certain  
 additional information that can be published by the individual collaborations
 regarding details of their analyses. We also mention those aspects where the 
 collaborations may adopt analysis practices that are more consistent with
 each other. We hope that these comments will be useful for future efforts. 

\section{Tevatron Observables}
\label{observables} 
\hspace*{0.18in}
We summarize here the observables described by CDF and D\O\ in their
 respective publications~\cite{cdfwmass,d0wmass,cdfwwidth,d0wwidth,cdfr,d0r1a,d0r1b}. 
The directly measured $W$ boson mass and 
 width~\cite{cdfwmass,d0wmass,cdfwwidth,d0wwidth} correspond to the
 pole mass $M_W$ and pole width $\Gamma_W$
 in the Breit-Wigner line shape with energy-dependent
 width, as defined by the differential cross section
\begin {eqnarray}
\frac {d\sigma} {d Q}={\cal L}_{q\bar{q}}(Q)
\frac {Q^2} {(Q^2-M_W^2)^2+{Q^4\Gamma_W^2/M_W^2}} \quad ,
\label{breitwigner}
\end {eqnarray}
 where $Q$ is the center-of-mass energy of the annihilating partons. 
${\cal L}_{q\bar{q}}(Q)$ 
represents the partonic luminosity  in hadron-hadron 
 collisions
\begin {eqnarray}
{\cal L}_{q\bar{q}}(Q) = {\frac{2Q}{s}} \sum_{i,j}\int_{Q^2/s}^1 
 {\frac{d x}{x}}
f_i(x,Q^2)f_j(Q^2/sx,Q^2)  \quad ,
\end {eqnarray}
where $i$ and $j$ represent parton flavors,  
 $f_{i,j}$ represent the respective parton distribution
 functions, $x$ is the momentum fraction of the parton,  and $\sqrt{s}$ 
 is the hadron-hadron center-of-mass energy. 

 The $W$ decay channels used for these 
 measurements~\cite{cdfwmass,d0wmass,cdfwwidth,d0wwidth} are the
 $e \nu$ channel (by CDF and D\O) and the $\mu \nu$
 channel (by CDF). 
The $W$ boson mass and width are extracted by analyzing the Jacobian
edge and the high mass tail respectively of the transverse mass
 ($m_T$) distribution 
\begin{eqnarray}
m_T (l^\pm \nu) = \sqrt{2 \; p_T(l^\pm) \; p_T(\nu)
  \left(1-\cos\left(\phi(l^\pm)-\phi(\nu)\right)\right)} \quad ,
\label{eq:mtdefn}
\end{eqnarray}
where $p_T$ and $\phi$ represent the transverse momentum and azimuthal angle
 respectively of the leptons.  
D\O\ has also measured the $W$ boson mass by analyzing the Jacobian edge
 in the electron and neutrino $p_T$ distributions. 
The CDF result for the $W$ boson mass is quoted using the $m_T$ fit, while
 the D\O\ result combines the $m_T$ fit and the lepton $p_T$ fits taking
 the correlations into account. 
 
The $W$ boson width is also extracted~\cite{cdfr,d0r1a,d0r1b}
  from the measured ratio of
 partial cross sections
\begin{eqnarray}
R & \equiv & \frac {\sigma_W \cdot B (W \rightarrow e \nu)}
{\sigma_Z \cdot B (Z \rightarrow e e)} \nonumber \\
 & = & \frac{\sigma_W}{\sigma_Z} \frac{\Gamma_Z}{\Gamma(Z \rightarrow ee)
} \frac{\Gamma(W \rightarrow e \nu)}{\Gamma_W}
\label{reqn}
\end{eqnarray}
by using as inputs the calculated ratio of total cross sections, 
 the measured $Z \rightarrow ee$ branching ratio from LEP and
 the SM calculation of the partial width
 $\Gamma(W \rightarrow e \nu)$. 

 Eqn.~\ref{breitwigner} gives the differential cross section for the 
 $W$ Drell-Yan process. The extraction of $M_W$ and $\Gamma_W$ 
 using Eqn.~\ref{breitwigner} 
 assumes the following:
\begin{enumerate}
\item the $W$ boson propagator can be described by the relativistic
 Breit-Wigner distribution in quantum field theory, and
\item the production of $W$ bosons in hadron-hadron collisions
 can be described by a factorizable process, with no additional interactions
 between the initial and final states. Since the leptonic final states
 are used for the measurements, there is no strong interaction between 
 the initial state and the final state. Electroweak corrections are considered
 in the analyses. Higher twist effects, which in principle alter the
 effective partonic luminosity factor ${\cal L}_{q\bar{q}}(Q)$, are in 
 practice negligible for $Q^2 = M_W^2$. 
\item backgrounds to the $W$ Drell-Yan process from $WW$, $WZ$ and $t \bar{t}$ 
 production are small; in practice, these rare processes are further suppressed
 by analysis selection cuts and produce essentially no contamination. 
\end{enumerate}

 In addition to the above, the extraction of $ B (W \rightarrow e \nu)$ and
  from $R$ assumes that the $W$ and $Z$ boson couplings to the leptons and light
 quarks are known, so that the inclusive cross section ratio 
 $\sigma_W/\sigma_Z$ can be calculated. The $Z$ boson leptonic branching
 ratio is well-measured at LEP. Uncertainties associated with
 $\sigma_W/\sigma_Z$ are discussed in Section~\ref{indirectwwidth}.
 
\section{$W$ Boson Mass}
\label{wmass}
\hspace*{0.18in}
The  Run 1 $W$ boson mass measurements  
from CDF~\cite{cdfwmass} and D\O\
 \cite{d0wmass} are
\begin{eqnarray}
M_W & = & 80.433 \pm 0.079 \;  \; {\rm GeV}  \;  \; {\rm (CDF)} \quad , \nonumber \\
M_W & = & 80.483 \pm 0.084 \;  \; {\rm GeV}  \;  \; {\rm (D\O)} \quad .
\end{eqnarray}
 We discuss the sources of uncertainty and classify them as being
 either uncorrelated between the two experimental results, or (partially or
 completely) correlated. 
\subsection{Uncorrelated Uncertainties}
\hspace*{0.18in} 
 The measurement and analysis techniques used by both experiments rely
 extensively on internal calibration and collider data to measure detector
 response and constrain theoretical model inputs. The bulk of the uncertainty
 is therefore uncorrelated. We itemize the uncorrelated sources below. 
 The following discussion also applies to the uncorrelated uncertainties
 in the direct measurement of the $W$ boson width (see Section~\ref{directwwidth}). 
\begin{itemize}
\item $W$ statistics in the kinematic distributions used for the mass
 fits. 
\item Detector 
 energy response and resolution measured using resonances ($Z$, $J/\psi$, 
 $\Upsilon$ and $\pi^0$). Model uncertainty from resonance line shapes
 is negligible. These data are used for the calibration of lepton
 energy response (calorimetry and tracking for electrons
 and tracking for muons). The $Z$ data are also used for calibrating the
 calorimeter response to the hadronic activity recoiling
 against the vector boson. In the CDF analysis, the lepton response and
 resolution and the hadronic recoil are modelled by empirical functions
 whose parameters are constrained independently
 for the electron and muon channel. Therefore in the internal
 CDF combination of these measurements, uncertainties
 in the lepton and recoil models are  uncorrelated between
 channels. D\O\ performs independent empirical fits to their data
 which are uncorrelated with CDF fits.  
\item Selection biases and backgrounds are unique to each experiment and
 are measured mostly from collider data, with some input from detector
 simulation for estimating selection bias. 
  These uncertainties are uncorrelated between the CDF
 electron and muon channel measurements. CDF has no selection bias
 for electrons (in contrast with D\O) 
 because the selection cuts  rely more heavily on tracking rather
 than calorimeter isolation, and because of a more inclusive $W \rightarrow e \nu$
 triggering scheme.  
\item The distribution of the transverse momentum ($p_T$)
 of the $W$ boson is a model input, which
 each experiment constrains individually by fitting the $Z$ boson $p_T$
 distribution. Phenomenological models such as that of Ellis, Ross and Veseli~\cite{ellis}
 or that of Ladinsky and Yuan~\cite{ly} are treated as 
 empirical functions which, after folding in the detector response, 
 adequately describe
 the observed $p_T(Z)$ distribution. The $p_T$ distribution is specified by
 model parameters along with $\Lambda_{QCD}$ and the parton distribution 
 functions (PDFs). The uncertainty is dominated by $Z$ statistics, with small
 dependence on the PDFs and $\Lambda_{QCD}$. The latter introduces a small
 correlation between the two experiments which can be neglected at this 
 level\footnote{Individual
 sources of uncertainty below about 3 MeV are typically not enumerated by the CDF and
 D\O\ experiments when the results are reported.}. 
 A potentially correlated uncertainty in the theoretical relationship between
 the $W$ boson and the $Z$ boson $p_T$ spectra is assumed to be negligible. 
 There is a small (3 MeV) correlated component in the $p_T(W)$ uncertainty 
 between the CDF electron and
 muon channel results. 
\item The  sources of background are $Z \rightarrow ll$ where
 one of the leptons is lost, 
 $W \rightarrow \tau \nu \rightarrow e/\mu + \nu \bar{\nu} \nu$, and 
 misidentified QCD jet events. The $Z \rightarrow ll$ background is
 estimated using individual detector simulations. The uncertainty on the
 $W \rightarrow \tau \nu \rightarrow e/\mu + \nu \bar{\nu} \nu$ background is 
 negligible. The jet misidentification background is estimated by using
 loosely defined lepton data samples which enhances the background
 contribution (D\O), or by selecting lepton candidates that fail quality cuts
 (CDF). While the techniques are similar in principle they differ in detail. 
 CDF has also confirmed the jet misidentification background estimate using
 a photon conversion sample. The background uncertainties and cross-checks
 are statistics-limited and therefore independent. 
\end{itemize}
 Table~\ref{massucerrors} shows the contributions to the uncertainty which
 are uncorrelated between the CDF and D\O\ measurements, taken from the
 respective 
 publications~\cite{cdfwmass,d0wmass}
 of the 1994-95 data as examples. All of these 
 uncertainties should reduce in the future with more data, as the detector
 simulation and  production/decay model is tuned with higher precision, 
 and backgrounds are reduced with tighter cuts.

\begin{table}[hbtp]
\caption{Uncorrelated uncertainties (MeV) 
  in the CDF~\protect\cite{cdfwmass} and D\O\ \protect\cite{d0wmass}
 $W$ boson mass
 measurements from the 1994-95 (Run 1b) data. 
 $W$ boson decay channels used ($e$, $\mu$) are
listed separately.}
\medskip
\begin{tabular}{lccc} 
\hline
Source & CDF $\mu$ & CDF $e$ & D\O\ $e$ \\
\hline                     
$W$ statistics  & 100 & 65 &  60  \\
\hline                     
Lepton scale  & 85 & 75 &  56  \\
\hline                     
Lepton resolution & 20 & 25 &  19  \\
\hline                     
$p_T(W)$   & 20 & 15 &  15  \\
\hline                     
Recoil model & 35 & 37 &  35 \\
\hline                     
Selection bias  & 18 & - &  12  \\
\hline                     
Backgrounds & 25 & 5 &  9  \\
\hline                     
\end{tabular}
\label{massucerrors}
\end{table}

\subsection{Correlated Uncertainties}
\hspace*{0.18in} 
Sources of correlated uncertainty are associated with the modeling of
 $W$ production and decay, which we itemize below. The uncertainties 
 are  fully correlated between CDF and D\O, with possibly
 different magnitudes. 
\begin{itemize}
\item The $W$ boson kinematic
 distributions used in the fits are invariant under longitudinal boosts
 because they are derived from transverse quantities. The sensitivity to the
 PDFs arises because of acceptance cuts on the charged
 lepton rapidity. As the rapidity acceptance increases the sensitivity to PDFs
 reduces. The D\O\ $W$ boson mass measurement 
 includes electrons up to pseudorapidity $| \eta | <2.5$, 
 and the CDF measurement includes electrons and muons up to $| \eta | < 1.0$. The PDF 
 uncertainty is correlated but different for the two measurements. 
\item The Breit-Wigner line shape  has an uncertainty due to the
 variation in the mass dependence of the partonic
  luminosity. This
 is a small contribution which D\O\ quotes separately, but CDF subsumes
 into the overall
 PDF uncertainty. 
\item QED radiative corrections in leptonic $W$ boson decays
 are evaluated by both experiments using the Berends and Kleiss~\cite{berends} 
 calculation. The uncertainty is evaluated by comparing to the PHOTOS~\cite{photos} program and/or the 
 calculation of Baur {\em et al.}~\cite{baur}.
 The higher order QED effects have a different impact on the electron and muon channel
 measurements from CDF and the electron measurement from D\O\ due to differences in energy measurement
 techniques. We find that in the combined electron and muon channel result of CDF, the effective uncertainty
 due to QED radiative corrections is 11 MeV. This contribution  is fully correlated with the
 corresponding uncertainty in the D\O\ result. 
\item The $W$ width 
 input into the $W$ boson mass measurement is provided  differently by CDF
 and D\O. CDF uses the SM prediction 
 for $\Gamma_W$ for the fitted value of
 $M_W$ and the resulting uncertainty is negligible. D\O\ uses the 
 indirect measurement of the
 $W$ width which is extracted from the D\O\ measurement of the ratio
 $\sigma(W \rightarrow e \nu) / \sigma(Z \rightarrow e e) $. Since the
 line-shape fits performed by CDF and D\O\ 
for the $W$ mass are sensitive to the assumed $W$ width, we
 require that both experimental results use a consistent treatment of the
 uncertainty associated with the width input. For the purpose of 
 combining the results, we take the 10 MeV uncertainty quoted by D\O\ 
 to be the correlated error. 
\end{itemize}
Table~\ref{masserrors} shows the correlated systematic uncertainties, taken
 from~\cite{cdfwmass} and~\cite{d0wmass} respectively.

\begin{table}[hbtp]
\caption{Systematic uncertainties (MeV) 
 from correlated sources in the $W$ boson mass
 measurements~\protect\cite{cdfwmass,d0wmass}. }
\medskip
\begin{tabular}{lcc} 
\hline                     
Source & CDF & D\O\ \\
\hline                     
PDF \& parton luminosity & 15 &  7 $\oplus$ 4  \\
\hline                     
Radiative Corrections & 11 & 12  \\
\hline                     
$\Gamma_W$ & 10 &  10  \\
\hline                     
\end{tabular}
\label{masserrors}
\end{table}

\subsection{Combination of Results}
\label{matrix}
\hspace*{0.18in} 
 We use the Best Linear Unbiased Estimate~\cite{blue} method,
  which is also used in~\cite{d0wmass},
 to construct the covariance matrix between the CDF and D\O\
 measurements. For each source of correlated error, we construct a 
 2-component vector $\delta_i \vec{M}_W$ whose components are the individual
 uncertainties quoted in Table~\ref{masserrors}, $i.e.$  
 $\delta_i \vec{M}_W =  (\delta_i M_W^{\rm CDF}, 
 \delta_i M_W^{\rm D\O})$ for the
 $i^{\rm th}$ source of uncertainty.   
 The contribution to the
 covariance matrix  from each source is given by 
 $V_i =  \delta_i \vec{M}_W
 ( \delta_i \vec{M}_W ) ^T$, where $T$ indicates the transpose. 
 The various sources of error are assumed to be uncorrelated with each other,
 hence we add the individual covariance matrices $V_i$ to obtain
 $V = \sum_i V_i$. This procedure gives
 us the off-diagonal term in the total covariance matrix $V$. 
 The diagonal terms
 are obtained from the square of each measurement's total error. 
 The square root of the off-diagonal covariance matrix element $\sqrt{V_{12}}$
 gives the
 total correlated error between the CDF and D\O\ measurements of 19 MeV.
 The correlation coefficient, defined by $V_{12} / \sqrt{V_{11} V_{23}}$, is 
 $ 19^2 / (79 \times 84) = 0.054$. 
  
The combined $W$ mass $M_W$ for the set of two $W$ mass measurements $m_i$
 and their covariance matrix $V$ is given by
\begin{equation}
   M_W = (\sum_{i,j=1}^{2}{H_{ij}}\,\,m_j)
  \, / \,(\,\sum_{i,j=1}^{2}{H_{ij}}\,) \quad  ,
\end{equation}
 where $H \equiv V^{-1}$ and $i,j$ run over the two $W$ mass measurements being
 combined. 
The combined error is given by
\begin{equation}
 \sigma(M_W) = (\,\sum_{i,j=1}^{2}{H_{ij}}\,)^{-1/2} \quad ,
\end{equation}
and the $\chi^2$ for the combination is given by
\begin{equation}
  \chi^2 = \sum_{i,j=1}^{2}{(m_i - M_W) \,\, H_{ij} \,\, (m_j - M_W )} \quad  .
\end{equation}

 Using this procedure, we obtain the combined result for the Tevatron collider
 \begin{equation}
 M_W^{\rm Tevatron} = 80.456 \pm 0.059 \; \; {\rm GeV} \quad ,
\label{tevatronmw}
 \end{equation}
 with $\chi^2 = 0.2$ and probability of 66\%. 

 We note that the various $W$ mass measurements from D\O\ 
 are internally combined
 by D\O\ \cite{d0wmass}
using the same technique that we describe above. CDF combines its
 internal measurements~\cite{cdfwmass}
  using a slightly different formulation, where
 the measurements are combined using only the uncorrelated errors, and then
 the correlated errors are added in quadrature. When the correlated errors
 are small with positive correlation coefficients, 
 as we have here, the two formulations give very similar results. 

 The result of Eqn.~\ref{tevatronmw} is not very different from a simple 
 average ignoring all correlations ($80.456 \pm 0.057$ GeV). This is due to the uncertainties being
 dominated by the uncorrelated components. As mentioned before, the uncorrelated
 sources of uncertainty will reduce with higher statistics. Therefore
 correlated theoretical errors such as QED radiative corrections may dominate
 in the future, in which case the error analysis we have presented here
 becomes more important. 

 The combination of the Tevatron collider average 
 with the UA2 measurement~\cite{UA2} of
 \begin{equation}
 M_W^{\rm UA2} = 80.36 \pm 0.37 \; \; {\rm GeV} \quad 
 \end{equation}
 with a common uncertainty of 19 MeV yields
 \begin{equation}
 M_W^{p\bar{p}} = 80.454 \pm 0.059 \; \; {\rm GeV} \quad . 
 \end{equation}
 Here we have taken the correlated component of the uncertainty
 between CDF and D\O\ as being fully correlated with the UA2 result, since
 all three hadron collider measurements are sensitive to the 
 PDFs, QED radiative corrections and $W$ width in much the same way. 

\begin{figure}[htb]
\vspace{-0.0cm}
\epsfxsize=3.5in
\centerline{\epsfbox{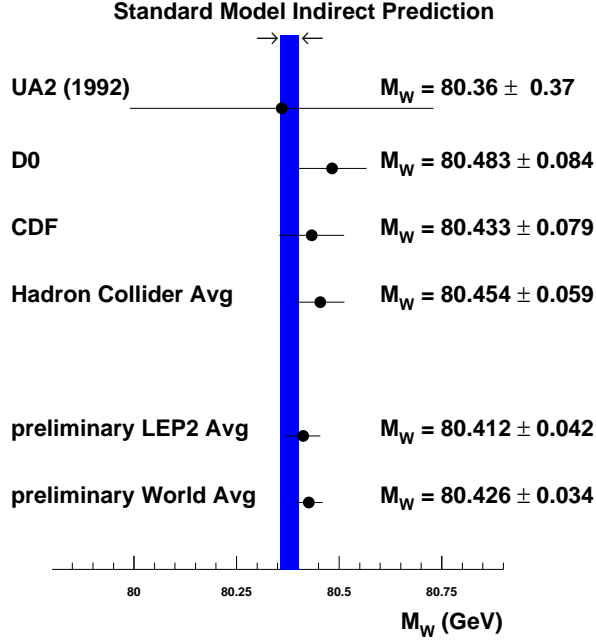}}
\vspace{0.5cm}
\caption {Direct measurements of the $W$ boson mass compared with 
 the SM prediction~\protect\cite{LEP} based on a fit to all $Z$-pole data
 and the direct top mass measurements.}
\label{wmassaverage}
\end{figure}

 Further combination with the preliminary LEP average~\cite{LEP} of
 \begin{equation}
 M_W^{\rm LEP} = 80.412 \pm 0.042 \; \; {\rm GeV} \quad 
 \end{equation}
 assuming no correlated uncertainty gives
 \begin{equation}
 M_W^{\rm world} = 80.426 \pm 0.034 \; \; {\rm GeV} \quad 
 \end{equation}
 as the preliminary world average (with $\chi^2=0.34$ and 56\% probability). 
 Figure~\ref{wmassaverage} shows the $W$ boson mass results,
  compared with the indirect value of $ 80.380 \pm 
 0.023$ GeV. The latter is obtained from a fit to all $Z$-pole data and the
 direct top mass measurements~\cite{LEP}, 
 as interpreted in the context of the SM. 

\section{$W$ Boson Width}
\label{directwwidth}
\hspace*{0.18in} 
The direct measurement of the $W$ boson 
 width is made by analyzing $W$ boson candidate
 events  with transverse 
mass above the Jacobian peak, which occurs for $m_T \sim 80$ GeV. The fitting range extends roughly
 between 100 GeV and 200 GeV, where the resolution effects from the 
 Jacobian peak are small. The $W$ boson 
 width analysis shares most of the issues of $W$ production and decay 
 modeling and the detector response
 with the $W$ boson mass analysis, and the sources of uncertainty are
 therefore similar. 

  As with the
 $W$ boson mass analysis, the model parameters are constrained by analysis 
 of internal data by each experiment
 separately. Therefore most of the uncertainties (shown in 
 Table~\ref{widthucerrors} for the 1994-95 data~\cite{cdfwwidth,d0wwidth}
 as examples)
 are uncorrelated. These uncertainties are
 also uncorrelated between the CDF electron and muon channel results. 
\begin{table}[hbtp]
\caption{Uncorrelated uncertainties (MeV) 
  in the CDF~\protect\cite{cdfwwidth} and D\O\ 
\protect\cite{d0wwidth}
 $W$ boson width measurements  from the 1994-95 (Run 1b) 
 data.
$W$ boson decay channels used ($e$, $\mu$) are
listed separately. }
\medskip
\begin{tabular}{lccc} 
\hline                     
Source & CDF $\mu$ & CDF $e$ & D\O\ $e$ \\
\hline                     
$W$ statistics  & 195 & 125 &  142  \\
\hline                     
Lepton energy scale  & 15 & 20 &  42  \\
\hline                     
Lepton $E$ or $p_T$ non-linearity & 5 & 60 &  -  \\
\hline                     
Recoil model  & 90 & 60 &  59  \\
\hline                     
$p_T(W)$ & 70 & 55 &  12 \\
\hline                     
Backgrounds & 50 & 30 &  42 \\
\hline                     
Detector modeling, lepton ID & 40 & 30 &  10  \\
\hline                     
Lepton resolution & 20 & 10 &  27  \\
\hline                     
Parton luminosity slope & - & - &  28  \\
\hline                     
\end{tabular}
\label{widthucerrors}
\end{table}

 The
  correlated sources of
 uncertainty are
\begin{itemize}
\item Parton distribution functions - the CDF and D\O\ analyses used
 different sets of PDFs to evaluate this uncertainty and quote 
 different contributions. The $W$ boson acceptance is similar
 in the direct measurements of the $W$ boson width since
 both experiments require lepton $p_T > 20$ GeV and $| \eta | < 1$. 
\item $W$ boson mass
\item QED radiative corrections
\end{itemize}
The Run 1 direct $W$ boson
 width measurements from CDF~\cite{cdfwwidth} and D\O\
 \cite{d0wwidth} are
\begin{eqnarray}
\Gamma_W & = & 2.05 \pm 0.13 \;  \; {\rm GeV}  \;  \; {\rm (CDF)} \quad , \nonumber \\
\Gamma_W & = & 2.231^{+0.175}_{-0.170} \;  \; {\rm GeV}  \;  \; {\rm (D\O) } \quad , 
\end{eqnarray}
where the total uncertainty is quoted. 
The correlated uncertainties for the 
 two measurements are shown in Table~\ref{widtherrors}.
\begin{table}[hbtp]
\caption{Systematic uncertainties (MeV) 
 from correlated sources in the direct $W$ boson width
 measurements~\protect\cite{cdfwwidth,d0wwidth}. }
\medskip
\begin{tabular}{lcc} 
\hline                     
Source & CDF & D\O\ \\
\hline                     
PDF & 15 &  27  \\
\hline                     
Radiative Corrections & 10 & 10  \\
\hline                     
$W$ boson mass & 10 & 15  \\
\hline                     
\end{tabular}
\label{widtherrors}
\end{table}
 The likelihood fit returns a slightly asymmetric statistical error for the D\O\ result. We symmetrize
 it by taking the arithmetic average and combine in quadrature with the total systematic uncertainty
 to obtain a total uncertainty of 173 MeV for the D\O\ result.  
 We use the procedure described in Section~\ref{matrix} to construct the covariance matrix, and use it
 to obtain the combined result
\begin{equation}
\Gamma_W^{\rm Tevatron} = 2.115 \pm 0.105 \; \; {\rm GeV} \quad ,
\label{tevatrongw}
\end{equation}
 with $\chi^2 = 0.7$ and probability of 40\%.
 The
 square root of the off-diagonal covariance matrix element gives the
 total correlated error of 26 MeV and a correlation coefficient
 of 0.03. As in the case of the $W$ mass combination, the
 uncorrelated errors dominate with the current statistics, and
 ignoring the correlation would produce a similar result ($2.115 \pm
 0.104$ GeV). However, in Run 2 at the Tevatron, which is expected to
 increase the statistics by a factor of $\sim20$, the
 correlated uncertainties on the theoretical inputs may dominate. 

\begin{figure}[htb]
\vspace{-0.0cm}
\epsfxsize=3.5in
\centerline{\epsfbox{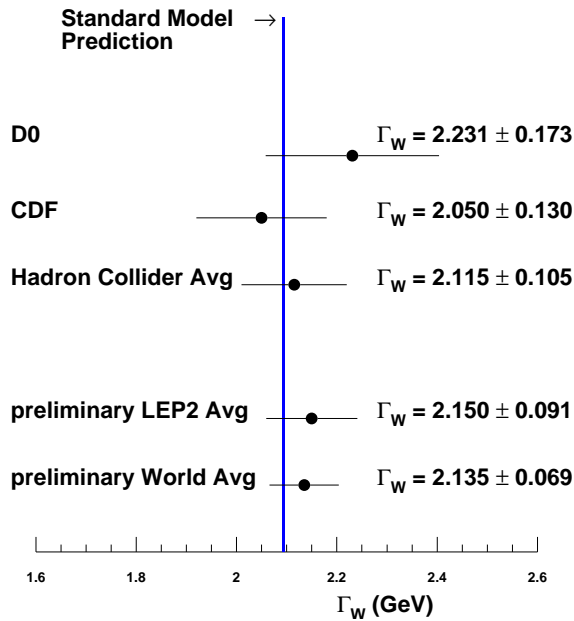}}
\vspace{0.5cm}
\caption {Direct measurements of the $W$ boson width compared with 
 the SM prediction~\protect\cite{PDG}.}
\label{figwwidth}
\end{figure}

 Combination of the Tevatron average 
 with the preliminary LEP average~\cite{LEP} of
 \begin{equation}
 \Gamma_W^{\rm LEP} = 2.150 \pm 0.091 \; \; {\rm GeV} \quad 
\label{lepwwidth}
 \end{equation}
 assuming no correlated uncertainty gives
 \begin{equation}
 \Gamma_W^{\rm world} = 2.135 \pm 0.069 \; \; {\rm GeV} \quad 
 \end{equation}
 as the preliminary world average (with $\chi^2=0.063$). 
 Figure~\ref{figwwidth} shows the $W$ boson width results,
  compared with the SM prediction of $ 2.0927 \pm 
 0.0025$ GeV~\cite{PDG}. 

\section{Joint Analysis of $W$ Boson Mass and Width}
\label{joint}
\hspace*{0.18in} 
 In this section we describe 
 the analysis for the joint direct measurement of the $W$ boson
 mass and the width. We do not allow external constraints on the mass
 and width parameters: 
 instead we propagate the uncertainties on the direct observables to the
  uncertainties on the 
 extracted Breit-Wigner parameters. This procedure will give us the values and 
 uncertainties 
 on $M_W$ and $\Gamma_W$  extracted from ``Tevatron data only'', as 
 well as their covariance.  
  
 We introduce the following terminology to 
 distinguish between the observables called $M_W$ and $\Gamma_W$ 
(which are returned by the fits to the data
 spectra) and the Breit-Wigner parameters of the same names (which we want to extract).
  We define the vector of observables $\vec{o} = (M_W^o, \Gamma_W^o)$
 and the vector of Breit-Wigner parameters $\vec{t} = (M_W^t, \Gamma_W^t)$. We approximate the
 functional dependence $\vec{o} \; ( \; \vec{t} \; ) $ by a linear dependence, so that 
  $\vec{o}$ and $\vec{t}$
 are related by a linear transformation. For the error analysis, we are interested in transforming
 the variations in $\vec{o}$ to variations in $\vec{t}$. This transformation is given by the matrix of
 derivatives $\Delta \equiv \partial \vec{o} / \partial \vec{t}$, such that 
\begin{equation}
\delta \vec{o} = \Delta \; \delta \vec{t} \quad .
\label{otdelta}
\end{equation}
The matrix of derivatives $\Delta$ is defined as  
\begin{equation}
  \Delta  =  
\left( \begin{array} {cc} \frac{\partial M_W^o}{\partial M_W^t} & \frac{\partial M_W^o}{\partial \Gamma_W^t} \\ \\
\frac{\partial \Gamma_W^o}{\partial M_W^t} & \frac{\partial \Gamma_W^o}{\partial \Gamma_W^t} \end{array} \right)  \quad .
\end{equation}

 The values of the matrix elements of $\Delta$ have been published by the CDF and D\O\ collaborations,
 using their Monte Carlo simulation programs~\cite{d0wmass,cdfwwidth,d0wwidth}. These simulation
 programs generate $W$ bosons according to the calculated mass, rapidity and $p_T$ distributions, 
 generate the decay products according to calculated angular distributions, and subject them to 
 parameterized detector response functions.
  The simulated decay leptons  are used to predict the distributions
 of the observables in the data.  

  The simulation and fitting programs demonstrate that the diagonal
 elements of $\Delta$ are unity. 
  The off-diagonal element $\frac{\partial M_W^o}{\partial \Gamma_W^t}$ is
 given by~\footnote{We use the value of the derivative quoted by D0, since CDF does not
   quote it. We assume that the same derivative would apply for both
   experiments since the $W$ boson kinematics and experimental resolutions are
   similar.} the 10 MeV variation in observed $M_W$ due to a 60 MeV variation in
 $\Gamma_W^t$~\cite{d0wmass}. The off-diagonal element
 $\frac{\partial \Gamma_W^o}{\partial M_W^t}$ is given by the 
 mean variation of 13 MeV~\footnote{The uncertainty in $\Gamma_W$ due to $M_W$
 is quoted as 10 MeV and 15 MeV by CDF and D\O\ respectively, which are
 consistent with being equal given that both experiments round the quoted
 systematics to the nearest 5 MeV due to Monte Carlo statistics. The kinematics
 and acceptance for both experiments are very similar, hence we expect the
 true sensitivity to be the same, for which our best estimate is their
 average of 13 MeV.}
 in observed $\Gamma_W$ for a 39 MeV variation in $M_W^t$
 \cite{cdfwwidth,d0wwidth}. Thus $\Delta$ is reported by CDF and D\O\ to be 
\begin{eqnarray}
  \Delta  =  \left( \begin{array} {cc} 1 & 0.17 \\ 0.33 & 1 \end{array} \right)    \quad .
\end{eqnarray}

We invert Eqn.~\ref{otdelta} to obtain $ \Delta^{-1} \; \delta \vec{o} = \delta \vec{t} $
and take the expectation value of the product of each vector and its transpose
\begin{equation}
 \Delta^{-1} \; < \delta \vec{o} \; (\delta \vec{o})^T > \; (\Delta^{-1})^T 
 \: =  \: < \delta \vec{t} \; (\delta \vec{t})^T >  \quad ,
\label{mwgwcorr}
\end{equation}
where $T$ denotes the transpose and $<...>$ denotes the expectation value.
 The left-hand-side of Eqn.~\ref{mwgwcorr}
  contains the covariance matrix of the observables 
$ < \delta \vec{o} \; (\delta \vec{o})^T > $, 
  and we identify the right-hand-side
  with the covariance matrix of the extracted Breit-Wigner parameters. 
 
The diagonal elements of $ < \delta \vec{o} \; (\delta \vec{o})^T > $ 
 are given by the variances of the individual Tevatron averages of the direct $W$ 
 boson mass and width (see Eqns.~\ref{tevatronmw}
 and~\ref{tevatrongw}), excluding the error contribution to $M_W$ due to $\Gamma_W$  and vice-versa.
In order to evaluate the off-diagonal matrix element, we analyze the various contributions to the 
 respective variances. The observables are obtained
  from fits to disjoint data
  samples\footnote{The $W$ mass fits are performed with the data satisfying $m_T < 90$ GeV or
 lepton $p_T  < 50$ GeV, while the fits for the $W$ width are performed with data satisfying $m_T > 100$ GeV.
  },  so that their statistical uncertainties are uncorrelated. 
 However, the observed values of $M_W$ and $\Gamma_W$ depend on
 the same detector parameters (such as energy scales and resolutions) and the same 
 theoretical parameters (such as
 parton distribution functions and QED radiative corrections). Hence the uncertainties in these ``nuisance''
 parameters propagate into correlated uncertainties between the observables. 

 To compute the off-diagonal term, we evaluate
  the uncertainty contribution to the observed $M_W$ and
 $\Gamma_W$ due to each of these 
 nuisance parameters. 
 The following procedure is followed: (i) remove the respective contribution 
 from the CDF and D\O\ results separately, by setting each to zero,   
 (ii) recompute the total error on the CDF+D\O\ average, and (iii) take
 the difference in quadrature 
 between the original total error and the reduced total error. 
  Table~\ref{mwgwtable} shows the uncertainty contributions from each 
  source to the CDF+D\O\ averages. 

 In the above procedure, we have followed the same assumption that is made by
 CDF and D\O\ in their publications - that the sources of uncertainty
 listed in Table~\ref{mwgwtable} are mutually independent. 
 This is a valid assumption given the statistics of these data. The lepton
 energy scale and resolution are derived from the observed peak position
 and width of the $Z$ boson mass distribution, which are essentially 
 decoupled. The $p_T(W)$ uncertainty is dominated by the statistical error
 of the $p_T(Z \rightarrow ll)$ measurement, although this may change in
  the future. The recoil model is tuned using transverse momentum 
 balance in $p_T(Z \rightarrow ll)$ events, where the lepton resolution
 is a small effect compared to the recoil resolution. 
  
\begin{table}[hbtp]
\caption{Correlated uncertainties (MeV) 
  between the CDF+D\O\ averages of 
 $M_W$ and direct $\Gamma_W$, due to nuisance parameters.}
\medskip
\begin{tabular}{lcc} 
\hline                     
Source & $M_W$ & $\Gamma_W$  \\
\hline                     
Lepton scale  &  37 & 17   \\
\hline                     
Lepton resolution & 12 & 11   \\
\hline                     
$p_T(W)$   & 9 & 24  \\
\hline                     
Recoil model & 20 & 35 \\
\hline                     
Detector modeling, selection bias  & 6 & 13   \\
\hline                     
QED radiative correction & 11 & 10 \\
\hline                     
\end{tabular}
\label{mwgwtable}
\end{table}

 We use the information from  Table~\ref{mwgwtable} to evaluate the covariance term
\begin{equation}
 < \delta M_W^o \; \delta \Gamma_W^o > \; = \sum_i \;  \delta_i M_W^o \;   \delta_i \Gamma_W^o    \quad ,
\label{deltamwgwcov}
\end{equation}
where the sum is performed over the various sources in  Table~\ref{mwgwtable}, and  
$\delta_i M_W^o$  and $ \delta_i \Gamma_W^o$ are the respective error contributions to  
 $M_W^o$ and $\Gamma_W^o$ from source $i$.  In this sum, the relative sign of each pair of
 factors $\delta_i M_W^o$  and $ \delta_i \Gamma_W^o$ determines the sign of the covariance
 contribution. 
The $W$ mass and width analyses were performed by each experiment 
 in a closely related manner, using the same
 simulation programs for both analyses. The 
  uncertainty contributions due to the nuisance parameters are completely
 correlated between the observed $M_W$ and $\Gamma_W$. Therefore 
  $\delta_i M_W^o$  and $ \delta_i \Gamma_W^o$ have the same sign in all 
 cases. To illustrate,
  in the cases of the lepton energy scale, lepton energy resolution,
 $p_T(W)$ and recoil modeling,  an
 increase in the respective
  parameter increases the observed values of both $M_W$ and
 $\Gamma_W$. Similarly, in the cases of detector modeling, selection bias
 and QED radiative correction,
 the bias in the shape of the $m_T$ or lepton $p_T$ spectrum affects both
 observables in the same direction. 

\begin{table}[hbtp]
\caption{Uncorrelated systematic uncertainties (MeV) 
  between the CDF+D\O\ averages of $M_W$ and direct $\Gamma_W$.}
\medskip
\begin{tabular}{lcc} 
\hline
Source & $M_W$ & $\Gamma_W$  \\
\hline                     
Backgrounds & 6 & 21 \\
\hline                     
PDF, parton luminosity & 12 & 22 \\
\hline                     
\end{tabular}
\label{mwgwtableuc}
\end{table}

 Table~\ref{mwgwtableuc} shows the systematic error contributions due to
 PDFs and backgrounds. We do not expect a strong correlation between the
 error contributions to the observed 
 mass and width from these sources, because the observables are
 derived from different ranges in $m_T$.
  Thus, in the case of the PDFs, a different $x$ range is relevant
 in each case. Furthermore, in the case of the $W$ mass, the uncertainty in the PDFs propagates
 mainly through acceptance effects, while in the case of the $W$ width, the main effect is through
 the relative normalization of the high and low $m_T$ regions. In the case of backgrounds,
 QCD jet misidentification produces the dominant background whose shape is 
 determined independently in the different $m_T$ regions. The
 sensitivity to the background shape and normalization is different in the fits for the mass and the width,
 since the shapes of the signal distributions are very different in the respective fitting windows. 
      On the basis of these arguments, 
 we take the contributions in Table~\ref{mwgwtableuc} to be  uncorrelated. They are not used directly
 in this joint error analysis; we present them here for completeness and future reference. 

 Evaluating Eqn.~\ref{deltamwgwcov}, we find  
 $< \delta M_W^o \; \delta \Gamma_W^o > \; =  43^2$ MeV$^2$, and
 the covariance matrix for $M_W^o$ and $\Gamma_W^o$ is 
\begin{eqnarray}
& & < \delta \vec{o} \; (\delta \vec{o})^T >  \: =  \:  \nonumber \\
& & \left( \begin{array} {cc} (59^2 - 10^2) \: {\rm MeV}^2 & 43^2 \: {\rm MeV}^2 \\
     43^2  \: {\rm MeV}^2 & (105^2-13^2)  \:  {\rm MeV}^2 \end{array} \right) 
 \quad ,
\end{eqnarray}
 where the removal of the 10 (13) MeV systematic on the individual measurement
 of $M_W$ ($\Gamma_W$) 
 due to $\Gamma_W$ ($M_W$) variation is shown explicitly.
 Substituting this result and $\Delta$ into Eqn.~\ref{mwgwcorr}  gives
 the covariance matrix 
 for the extracted Breit-Wigner parameters  $M_W^t$ and $\Gamma_W^t$ :
\begin{eqnarray}
< \delta \vec{t} \; (\delta \vec{t})^T >  \: =  \:
 \left( \begin{array} {cc} (59 \: {\rm MeV})^2  & -(33 \: {\rm MeV})^2 \\
        -(33 \: {\rm MeV})^2 & (106 \: {\rm MeV})^2 \end{array} \right)
  \quad .
\end{eqnarray}

 The negative sign of the covariance between $M_W^t$ and $\Gamma_W^t$ 
 can be understood as follows:  a higher value of the Breit-Wigner pole mass 
 increases the predicted number of events at high $m_T$, causing the
 inferred $\Gamma_W^t$ to reduce (given the number of observed events at
 high $m_T$). 
 Similarly, a higher value of the Breit-Wigner width increases
 the expected number of events on the high side of the Jacobian edge, 
 causing the inferred $M_W^t$ to reduce (given the observed position of
 the Jacobian edge). 

 We now describe the calculation of the $M_W$ and $\Gamma_W$ central values in
 the joint analysis. We
 shift the observed  value of each variable  by the corresponding
 difference of the other variable from its assumed value, 
 scaled by the appropriate partial derivative:
\begin{eqnarray}
M^t & = & M^o + b \; ( \Gamma_A - \Gamma^t )  \quad , \nonumber \\
\Gamma^t & = & \Gamma^o + a \; ( M_A - M^t )  \quad .
\label{jointequations}
\end{eqnarray}
 $M_A$ and $\Gamma_A$ denote
  the assumed values of the mass and width used in the
 width and mass analyses respectively. $\Gamma^o$ and $M^o$ are the values
 extracted in these individual analyses, and
  $a$ and $b$ denote the partial derivatives $a = \partial \Gamma / \partial
 M = 0.33$ and $b = \partial M / \partial \Gamma = 0.17$. Solving 
 these simultaneous linear equations for $M^t$ and $\Gamma^t$, we obtain the
 central values in the joint analysis. 

 CDF and D\O\ used different assumed values of the $W$ mass and width in
 their respective analyses. 
   For our simultaneous $M_W - \Gamma_W$ analysis, we need
 the individual Tevatron averages of $M_W$ and $\Gamma_W$ for which the inputs are 
 quoted using the same reference values of $\Gamma_W$ and $M_W$ respectively.
Thus we cannot use the results of  Eqn.~\ref{tevatrongw} and Eqn.~\ref{tevatronmw}
 for $\Gamma^o$ and $M^o$ directly.  
To arrive at the appropriate averages, we  use the partial derivatives mentioned above
 to ``shift'' the CDF and D\O\ measurements
 to common reference values of $M_A = 80.413$ GeV and $\Gamma_A = 2.080$ GeV. 
 
 This reference point is calculated as follows. In the width
 analysis, D\O\ assumed a $W$ mass value of 80.436 GeV while CDF assumed
 a value of 80.400 GeV. 
 We use the weights derived for combining the
 CDF and D\O\ width measurements (Sec.~\ref{directwwidth})
  to obtain the average $M_A = 80.413$
 GeV. Similarly, in the mass analysis, D\O\ assumed a $W$ width value
 of 2.062 GeV and CDF assumed a value of 2.096 GeV. Using the weights 
 derived for combining the mass measurements (Sec.~\ref{wmass}),
  we obtain the average $\Gamma_A = 2.080$ GeV. 

 For these coordinates of the reference point, the CDF and D\O\ 
 $W$ mass measurements shift by about 3 MeV each, and the $W$ width
 measurements shift by about 4 MeV and 8 MeV respectively for CDF and D\O. 
 The combination of these shifted values is then repeated according to the
 procedure described in Sec.~\ref{wmass} and~\ref{directwwidth}. We obtain
 the new individual Tevatron averages of 
  $\Gamma^o = 2.115$ GeV   and $M^o = 80.456$ GeV. These values are
 identical to those quoted in Eqn.~\ref{tevatrongw} and Eqn.~\ref{tevatronmw}, 
 proving that our calculated reference point is consistent with the original
 choices made by CDF and D\O. 

 We can now solve the simultaneous linear equations given in
 Eqns.~\ref{jointequations}, to obtain
\begin{eqnarray}
M_W^{\rm Tevatron} & = & 80.452 \pm 0.059 \; \; {\rm GeV}  \quad , \nonumber \\
\Gamma_W^{\rm Tevatron} & = & 2.102 \pm 0.106 \; \; {\rm GeV}  \quad , 
\label{jointvalues}
\end{eqnarray}
as the Tevatron results of the joint analysis. The correlation coefficient is $-0.174$.

 Finally, it is of interest for future, higher precision measurements
 of $M_W$ and $\Gamma_W$ to pursue this joint analysis technique. We expect
 most error contributions to scale with the statistics of the data. Assumptions
 that are made in providing external input for $\Gamma_W$ in the
  $M_W$ analysis are not necessary in this joint analysis technique. We also
 note that there is almost no loss of precision compared to the individual
 measurements. While this may seem surprising, the reason is the positive
 covariance induced between $M_W^o$ and $\Gamma_W^o$ by the uncertainties
 in the nuisance parameters. This means that an error in any of the nuisance
 parameters moves $M_W$ and $\Gamma_W$ in the same direction. But since
 an increase in one causes the other to reduce as mentioned above, this 
 overall negative feedback suppresses the systematic uncertainties from the
 nuisance parameters on both $M_W^t$ and $\Gamma_W^t$. This reduction
 in other systematic errors compensates for the information lost in excluding
 external mass and width input. 
\section{Indirect $W$ Boson Width and Leptonic Width}
\label{indirectwwidth}
\hspace*{0.18in} 
 The CDF and D\O\ measurements of $R$ (see Eqn.~\ref{reqn}) have been
 presented~\cite{cdfr,d0r1a,d0r1b} elsewhere. 
 We describe here the combination of
 the $R$ measurements and the extraction of $\Gamma_W$ from $R$ assuming
 the validity of the SM. We also combine this extracted
 value of $\Gamma_W$ with the directly measured $\Gamma_W$ from the 
 $m_T$ spectrum shape.
\subsection{Combination of $R$ Measurements}
\hspace*{0.18in} 
The published CDF~\cite{cdfr} and D\O\ \cite{d0r1a,d0r1b} 
 measurements of $R$ in the electron channel are
\begin{eqnarray}
R & = & 10.90 \pm 0.32 \; {\rm (stat)} \pm 0.30 \; {\rm (syst)} \;  \; {\rm (CDF)} \quad , \nonumber \\
R & = & 10.82 \pm 0.41 \; {\rm (stat)} \pm 0.36 \; {\rm (syst)} \;  \;
 {\rm (D\O\ Run \; 1a) }  \quad , \nonumber \\
R & = & 10.43 \pm 0.15 \; {\rm (stat)} \pm 0.23 \; {\rm (syst)} \;  \;
 {\rm (D\O\ Run \; 1b) }  \quad  ,
\end{eqnarray}
where Run 1a refers to the 1992-93 data and Run 1b refers to the 1994-95 data.
 The uncertainties are summarized in Table~\ref{Rerrors}. 
\begin{table}[hbtp]
\caption{Fractional uncertainties (in \%) 
  in the CDF~\protect\cite{cdfr} and D\O\ \protect\cite{d0r1a,d0r1b}
 measurements of $R$. ``1a'' and ``1b'' refer to the 1992-93 
 and 1994-95 data respectively. The column labelled ``common'' indicates
 the correlated error, taken as common between the D\O\ Run 1a and Run 1b
 measurements. The 1a and 1b columns indicate (in some cases additional)
 uncorrelated errors for the D\O\ measurements. The last column indicates
 the error components that are correlated between the CDF and the combined
 D\O\ measurements. }
\medskip
\begin{tabular}{lccccc} 
\hline                     
Source & CDF & \multicolumn{3}{c}{D\O\ } & CDF \& D\O\ \\
       &     & 1b & common & 1a &  Correlated \\
\hline                     
PDF  & 1.1 &   & 0.3 &  & 0.3 \\
$M_W$  & 0.1 &  & 0.1 & 0.7 & 0.1 \\
Boson $p_T$  & 0.2 &  & 0.1 & 0.4 & \\
Energy Scale  & 0.4 &  0.7 & & 0.3 & \\
Recoil Response & 0.6 &  0.2 & & 0.6 & \\
Clustering Algorithm  & - &   & 0.2 &  & \\
Generator  & - &   & 0.3 & & \\
Electroweak Corrections  & 1.0 &  & 1.0 & & 1.0 \\
Backgrounds  & 1.5 &  1.7 & 0.1 & 2.3 & \\
Efficiencies  & 1.5 &  0.6 & & 1.9 & \\
NLO QCD            & 0.6 &  - & & - & \\
Drell-Yan          & 0.2 &  - & & - & \\
\hline                     
Total Systematic  & 2.8 &  2.2 & 1.1 & 3.3 & 1.0 \\
\hline                     
Statistical & 2.9 &  1.4 & - & 3.8 & - \\
\hline                     
Total & 4.0 &  2.7 & 1.1 & 5.1 & 1.0 \\
\hline                     
\end{tabular}
\label{Rerrors}
\end{table}

We combine the D\O\ results from Run 1a and Run 1b taking the systematics 
 due to choice of PDF (0.3\%), the uncertainty in $M_W$ (0.1\%), the
 uncertainty in the boson $p_T$ spectrum (0.1\%), clustering algorithm
 dependence (0.2\%), physics generator issues (0.3\%), electroweak 
 radiative corrections (1.0\%)~\cite{baur}
  and Drell-Yan background (0.1\%), as 
 correlated error components, to obtain a total correlated uncertainty of
 1.1\%. The result for the combined D\O\ measurement
 is
\begin{eqnarray}
R^{\rm D\O\ } & = & 10.50 \pm 0.23 \; ({\rm uncorrelated}) \pm 0.12 \;
  ({\rm correlated})
\nonumber \\
& = & 10.50 \pm 0.26  \quad   .
\end{eqnarray} 

This D\O\ result is then combined with the CDF measurement. In this 
 combination,
 the systematics due to the choice of PDF (0.3\%), the 
 uncertainty in $M_W$ (0.1\%) and 
 higher-order electroweak corrections (1.0\%) are treated as 
 correlated uncertainties, to obtain a total correlated
 uncertainty of 1.0\%. The average $R$ value is
\begin{eqnarray}
R^{\rm Tevatron} & = & 10.59 \pm 0.20 \; ({\rm uncorrelated}) \pm 0.11 \; ({\rm correlated})
\nonumber \\
& = & 10.59 \pm 0.23  \quad   .
\end{eqnarray} 

\subsection{Extraction of $W$ Boson Width}
\hspace*{0.18in} 
 In the extraction of $\Gamma_W$ from $R$, the $Z \rightarrow ee$ 
 branching ratio is taken from the PDG~\cite{PDG2002} to be ($3.363 \pm 0.004$)\%.
  The inclusive cross section ratio $\sigma_W / \sigma_Z$ is calculated
 at NNLO using Van Neerven {\em et al.}~\cite{vanneerven}, with the following inputs:
$M_W^{\rm Tevatron} = 80.456 \pm 0.059 $ GeV,  $M_Z = 91.187 $ GeV, 
 $\Gamma_W = 2.06 \pm 0.05$ GeV, $\Gamma_Z = 2.490 $ GeV, and 
 sin$^2 \theta_W = 0.23124 $~\cite{PDG98}.
The renormalization and factorization scales are set to the boson mass. 
The calculated value of the ratio of inclusive cross sections is found to be
\begin{equation}
\frac{\sigma_W}{\sigma_Z} = 3.360 \pm 0.051 \quad .
\end{equation}

The dominant
 uncertainties (quoted in parentheses) in the calculation of the cross section 
 ratio are 
 due to PDFs (0.45\%), 
 $M_W$ (0.09\%), factorization scale (0.12\%), renormalization scale (0.06\%) and
 sin$^2 \theta_W$ (1.43\%).  The uncertainties due to $M_Z$, $\Gamma_W$
 and $\Gamma_Z$ are negligible. The uncertainty due to any input is
 estimated by varying the input by $\pm 1 \sigma$ and taking half of the
 difference between the results. The uncertainties due to the renormalization
 and factorization scales are estimated by varying the scales high and low
 by a factor of two. The uncertainty due to electroweak corrections is
 estimated by taking different conventions for sin$^2 \theta_W$. For our
 central value we use sin$^2 \theta_{eff}$ from LEP, which gives an effective
 Born approximation and minimizes higher order corrections. The on-shell
 value of sin$^2 \theta_W = 1 - (M_W / M_Z)^2$, however, is equivalent at
 tree level, but gives a $Z$ boson production cross section which is about 
 1.4\% higher. We include this variation as a systematic uncertainty on 
 the calculated $\sigma_W / \sigma_Z$. 

The value of the $W \rightarrow e \nu$ branching ratio extracted from the
 combined CDF and D\O\ measurement of $R$ using Eqn.~\ref{reqn}  is 
\begin{equation}
B(W \rightarrow e \nu) = (10.61 \pm 0.28) \; \% \: ({\rm Tevatron}) \quad .
\label{wbranchingratio}
\end{equation}
For comparison, the SM value of the branching ratio~\cite{PDG2002}
 is
\begin{eqnarray}
B(W \rightarrow e \nu) & = & \frac{\Gamma (W \rightarrow e \nu) }{\Gamma (W)}  \nonumber \\
 & = &  [3 + 6 \{1 + \frac{\alpha_S}{\pi}
 + 1.409 \; ( \frac{\alpha_S}{\pi} )^2 
 - 12.77 \; ( \frac{\alpha_S}{\pi} )^3 \} ]^{-1} \nonumber \\
 & = & 0.10820 \pm 0.00007 \quad ,
\label{smwbr}
\end{eqnarray}
where we have used $\alpha_S(M_W)  = 0.1224 \pm 0.0028$~\footnote{
 This value of $\alpha_S(M_W)$ is obtained
 by evolving 
$\alpha_S(M_Z)  = 0.1200 \pm 0.0028$ (Eqn.~10.50 of~\cite{PDG2002})
  from $M_Z = 91.19$ GeV to $M_W = 80.45$ GeV
 using Eqn.~9.4 of~\cite{PDG2002}.}. 

 The SM calculation of the
 $W$ boson leptonic partial width 
 is given by~\cite{rosner}
\begin{eqnarray}
\Gamma (W \rightarrow e \nu) & = & \frac{G_\mu M_W^3}{6 \pi \sqrt{2}} ( 1 + \delta^{SM} )
 \nonumber \\
& = & 227.1 \pm 0.6 \; {\rm MeV} \quad ,
\label{smwleptonicbr}
\end{eqnarray}
where $G_\mu = (1.16637 \pm 0.00001) \times 10^{-5}$ 
 GeV$^{-2}$ is the muon decay constant~\cite{PDG2002},
 $\delta^{SM} = -0.0035 \pm 0.0017$ is the ``oblique'' correction to the tree level
 partial width~\cite{rosner}, and $M_W^{\rm Tevatron} = 80.456 \pm 0.059 $ GeV.
 Using this value of 
 $ \Gamma (W \rightarrow e \nu) $, 
 the extracted value of $\Gamma_W$ is
\begin{equation}
\Gamma_W^{\rm Tevatron} = 2.141 \pm 0.057 \; {\rm GeV}  \quad  .
\end{equation}
For comparison, the SM prediction is 
$ \Gamma_W^{\rm SM} = 2.099 \pm 0.006 \; {\rm GeV} $, using Eqns.~\ref{smwbr} 
 and~\ref{smwleptonicbr}. 

 A comparison of the width extracted from $R$ with the directly measured width
 (Eqns.~\ref{tevatrongw} and~\ref{jointvalues}) provides an interesting test of the SM, 
 since the two methods are quite different.  
 In the former measurement assumptions are made about boson couplings, whereas
 the latter makes use of kinematics. This test is one of the main goals
 of the Tevatron electroweak physics program in the future, 
 as the precision of both measurements
 improves with more data. 
  
 If one is willing to make all the SM assumptions mentioned in Sec.~\ref{observables}, 
 it is possible to combine the indirect and direct measurements of $\Gamma_W$. The result
 should be used with care; for instance, it may not be used in global fits where boson
 couplings are free parameters, or in analyses of data where new physics can affect the
 $W$ boson branching ratios. 

 Given these caveats, we discuss other aspects of combining the indirect
 and direct measurements of $\Gamma_W$.
 We consider the correlation  induced by theoretical inputs used
  in the respective analyses:   
\begin{itemize}
\item  PDF uncertainties: we conclude that there is no significant correlation
  because different
 aspects of PDFs are relevant for each analysis. For the direct measurement of the width
 the PDFs influence the mass dependence of the Breit-Wigner line shape at high mass. 
 For the extraction of the width from $R$, the PDF's influence the boson acceptance
 via their rapidity distributions. The total cross section ratio $\sigma_W/\sigma_Z$
  is affected by the $u(x)/d(x)$ ratio of PDFs. 
\item Electroweak corrections, factorization and 
 renormalization scales and sin$^2 \theta_W$  play no significant role in the
 direct $\Gamma_W$ measurement. 
\item We consider the correlation induced by variation in $M_W$. 
The uncertainties in $R$ and $\sigma_W / \sigma_Z$ due
 to $M_W$ variation are of the same magnitude and sign\footnote{With increasing $M_W$, $R$ 
 reduces due to increased acceptance, and $\sigma_W $ also reduces.} and therefore cancel 
 in $B(W \rightarrow e \nu)$. 
 The uncertainty in the SM calculation of
 $\Gamma (W \rightarrow e \nu)$ due to uncertainty in $M_W$ is 0.3\%, which is transferred
 to the extracted $\Gamma_W$ as a 7 MeV uncertainty. This is anti-correlated\footnote{
 With increasing $M_W$, the calculated $\Gamma (W \rightarrow e \nu)$ increases,
 whereas the directly measured $\Gamma_W$ decreases. } with 
 the corresponding $M_W$ uncertainty on the direct $\Gamma_W$ measurement (13 MeV). 
\end{itemize}

 Taking the anti-correlation induced by $M_W$ variation 
 into account\footnote{Ignoring the anti-correlation changes the result
 and uncertainty by less than 1 MeV.}, 
 we find the Tevatron combined (direct and indirect) result 
\begin{equation}
\Gamma_W^{\rm Tevatron} = 2.135 \pm 0.050 \; {\rm GeV}  \quad  .
\end{equation}
 The $\chi^2$ of this combination is 0.05 with a probability of 83~\%, indicating consistency
 between the direct and indirect measurements. 
Further combining with the preliminary LEP direct measurement (Eqn.~\ref{lepwwidth}) gives
\begin{equation}
\Gamma_W^{\rm world} = 2.139 \pm 0.044 \; {\rm GeV}  \quad  .
\end{equation}
Our world average differs from the PDG~\cite{PDG2002} value of 
 $\Gamma_W^{\rm world} = 2.118 \pm 0.042 \; {\rm GeV} $ because
 we have considered the correlations between the CDF and D\O\
 measurements, which were ignored in~\cite{PDG2002}.

\subsection{Extraction of $W$ Leptonic Width}
\hspace*{0.18in} 
We may use the extracted value of the $W \rightarrow e \nu$ branching
  ratio (Eqn.~\ref{wbranchingratio})
 and the directly measured  total $W$ width (Eqn.~\ref{tevatrongw})
 to obtain a measurement of
 the $W$ leptonic partial width
\begin{eqnarray}
\Gamma (W \rightarrow e \nu) & = & \Gamma_W \times 
 B(W \rightarrow e \nu) \nonumber \\
& = & 224 \pm 13 \; {\rm MeV} \: ({\rm Tevatron}) \quad .
\end{eqnarray} 
 The fractional 
 uncertainty in the direct $\Gamma_W$ (5.0\%) dominates over the fractional
 uncertainty in $B(W \rightarrow e \nu)$ (2.4\%). This measurement of 
 $\Gamma (W \rightarrow e \nu)$ is in
 good agreement with the SM calculation given in 
 Sec.~\ref{indirectwwidth}. 
\section{Suggestions for Future Publications}
\label{suggestions} 
\hspace*{0.18in}
There are a few instances where CDF and D\O\ have treated
 uncertainties differently in their respective analyses. For future
 efforts it would be helpful if a consistent treatment were adopted
 by both collaborations with mutual agreement. We itemize these
 cases below.
\begin{itemize}
\item The uncertainty due to PDFs has been treated differently in two 
 respects, the acceptance-related uncertainty and the parton luminosity
 uncertainty. 

 The acceptance-related effects were studied by varying
 PDFs, but these variations differed between the CDF and D\O\ analyses. 
 CDF used their $W \rightarrow e \nu$ charge asymmetry data 
 to constrain the PDF variation, whereas the variation considered by D\O\
 did not have this constraint because
 D\O\ did not have electron charge discrimination capability in Run 1.
 With the Run 2 detector D\O\ can also make this measurement. D\O\
 and CDF have also demonstrated that the boson (decay lepton) 
 rapidity distributions measured in $Z$ ($W$) boson events can
 provide additional PDF constraints, especially when forward
 lepton coverage is included. The optimal use of all this information
 would be to impose a combined constraint on PDFs using CDF and 
 D\O\ data, and then propagate the same  PDF uncertainty
 into their respective analyses.

 This approach may also be applied to the parton luminosity 
 uncertainty. Furthermore we suggest that the possibility of
 correlation between the parton luminosity uncertainty and the
 acceptance-related PDF uncertainty be studied. We suggest that
 these components be quoted separately along with their
 correlation. 
   
\item The treatment of the $W$ width input in the $W$ mass analysis
 and vice versa should be standardized. We suggest that both
 experiments adopt a common reference point based on available
 information. This also implies using a fixed width (mass) for the
 mass (width) fitting instead of building in a SM relationship
 between these parameters. This approach will facilitate the
 combination of the 1-parameter measurements, the 2-parameter
 joint analysis, and the comparison to theory.  
\end{itemize}
\section{Conclusion}
\hspace*{0.18in} 
We have presented the Run 1 results on the 
$W$ boson mass and width from the CDF and D\O\
 experiments, and examined their 
sources of uncertainty to identify the correlated components. We
 have used the covariance matrix technique to combine the respective 
 measurements from the two experiments. 
 The $\chi^2$ probability for each combination is good
 indicating that the measurements are consistent. We have also reported 
 the values and covariance matrix of the $W$ mass and direct $W$ width 
 measurements from their joint analysis. 
 Finally, we have combined the measurements of the ratio of $W$ and $Z$ boson
 cross sections, and extracted the combined value of the $W$ leptonic 
 branching ratio and the total $W$ width. The measurements of the
 $W$ width using the direct and indirect techniques are consistent, providing
 a test of the standard model. 
 We have also extracted the $W$ leptonic partial width from the measured
 total $W$ width and the leptonic branching ratio.   We have documented
 the methodologies that can provide the basis for future work with
 data of higher precision. 

\section*{Acknowledgements }
The CDF Collaboration thanks the Fermilab staff and the technical staffs 
of the participating institutions for their vital contributions.  
 This work was supported by the U.S. Department of Energy and National Science
 Foundation; the Italian Istituto Nazionale di Fisica Nucleare; the Ministry 
of Education, Culture, Sports, Science, and Technology of Japan; the Natural 
Sciences and Engineering Research Council of Canada; the National Science 
Council of the Republic of China; the Swiss National Science Foundation; the 
A.P. Sloan Foundation; the Bundesministerium fuer Bildung und Forschung, 
Germany; and the Korea Science and Engineering Foundation (KOSEF); the Korea 
Research Foundation; and the Comision Interministerial de Ciencia y 
Tecnologia, Spain.

%
The D\O\ Collaboration thanks the staffs at Fermilab and collaborating institutions, 
and acknowledge support from the 
Department of Energy and National Science Foundation (USA),  
Commissariat  \` a l'Energie Atomique and 
CNRS/Institut National de Physique Nucl\'eaire et 
de Physique des Particules (France), 
Ministry of Education and Science, Agency for Atomic 
   Energy and RF President Grants Program (Russia),
CAPES, CNPq, FAPERJ, FAPESP and FUNDUNESP (Brazil),
Departments of Atomic Energy and Science and Technology (India),
Colciencias (Colombia),
CONACyT (Mexico),
Ministry of Education and KOSEF (Korea),
CONICET and UBACyT (Argentina),
The Foundation for Fundamental Research on Matter (The Netherlands),
PPARC (United Kingdom),
Ministry of Education (Czech Republic),
A.P.~Sloan Foundation,
and the Research Corporation.

\end{document}